\documentclass[aps,prb,twocolumn,showpacs,amssymb,superscriptaddress,balancelastpage]{revtex4-1} 

\usepackage{amsmath,amssymb,amsfonts,amsthm}   
\usepackage{graphicx} 											
\usepackage{setspace}                           
\usepackage{hyperref}                                         
\usepackage{bm}
\usepackage{color}

\newcommand{\tr}{\textnormal{tr}}
\newcommand{\mathsym}[1]{{}}
\newcommand{\unicode}[1]{{}}

\begin{document}

\title{\noindent Photocurrent Response of Topological Insulator Surface States}
\author{Alexandra Junck}
\affiliation{\mbox{Dahlem Center for Complex Quantum Systems and Fachbereich Physik, Freie Universit\"at Berlin, 14195 Berlin, Germany}}
\author{Gil Refael}
\affiliation{\mbox{Department of Physics,
California Institute of Technology, 1200 E. California Blvd, MC114-36,
Pasadena, CA 91125 }}
\affiliation{\mbox{Dahlem Center for Complex Quantum Systems and Fachbereich Physik, Freie Universit\"at Berlin, 14195 Berlin, Germany}}
\author{Felix von Oppen}
\affiliation{\mbox{Dahlem Center for Complex Quantum Systems and Fachbereich Physik, Freie Universit\"at Berlin, 14195 Berlin, Germany}}
\date{\today}
\begin{abstract}
We study the photocurrent response of topological insulator surface states to circularly polarized light for arbitrary oblique incidence. We describe the surface states within a Dirac model, including several perturbations such as hexagonal warping, nonlinear corrections to the mode velocity, and applied magnetic fields. We find that the photogalvanic current is strongly suppressed for the usual orbital coupling, prompting us to include the weaker Zeeman coupling. We find that the helicity-independent photocurrent dominates over the helicity-dependent contributions.
\end{abstract}
\maketitle


{\em Introduction.---}Topological insulators (TI) recently emerged as a central theme in condensed matter physics.\cite{hasan2010} The intense interest in this new state of matter is rooted in their unique properties.\cite{kane2005,bernevig2006,konig2007,fu2007a,moore2007,hsieh2008} In addition to a band gap, TIs have conducting surface states with remarkable properties. They are protected against backscattering by time-reversal symmetry, and are helical: each surface-momentum state possesses a unique spin direction. The unique properties of the surface are responsible for their exotic electromagnetic properties,\cite{qi2008,essin2009} and might be used to realize topological superconducting phases hosting Majorana modes when brought into contact with s-wave superconductors.\cite{fu2008} Bismuth-based compounds were among the first materials predicted to be three-dimensional TIs,\cite{fu2007,zhang2009,yan2010} a prediction verified experimentally by angle-resolved photoemission\cite{hsieh2008,hsieh2009,chen2009,xia2009,zhang2009b,wang2011} and scanning tunneling spectroscopy.\cite{zhang2009b,zhanybek2010}
 
\begin{figure}[t]
 \includegraphics[width=8cm]{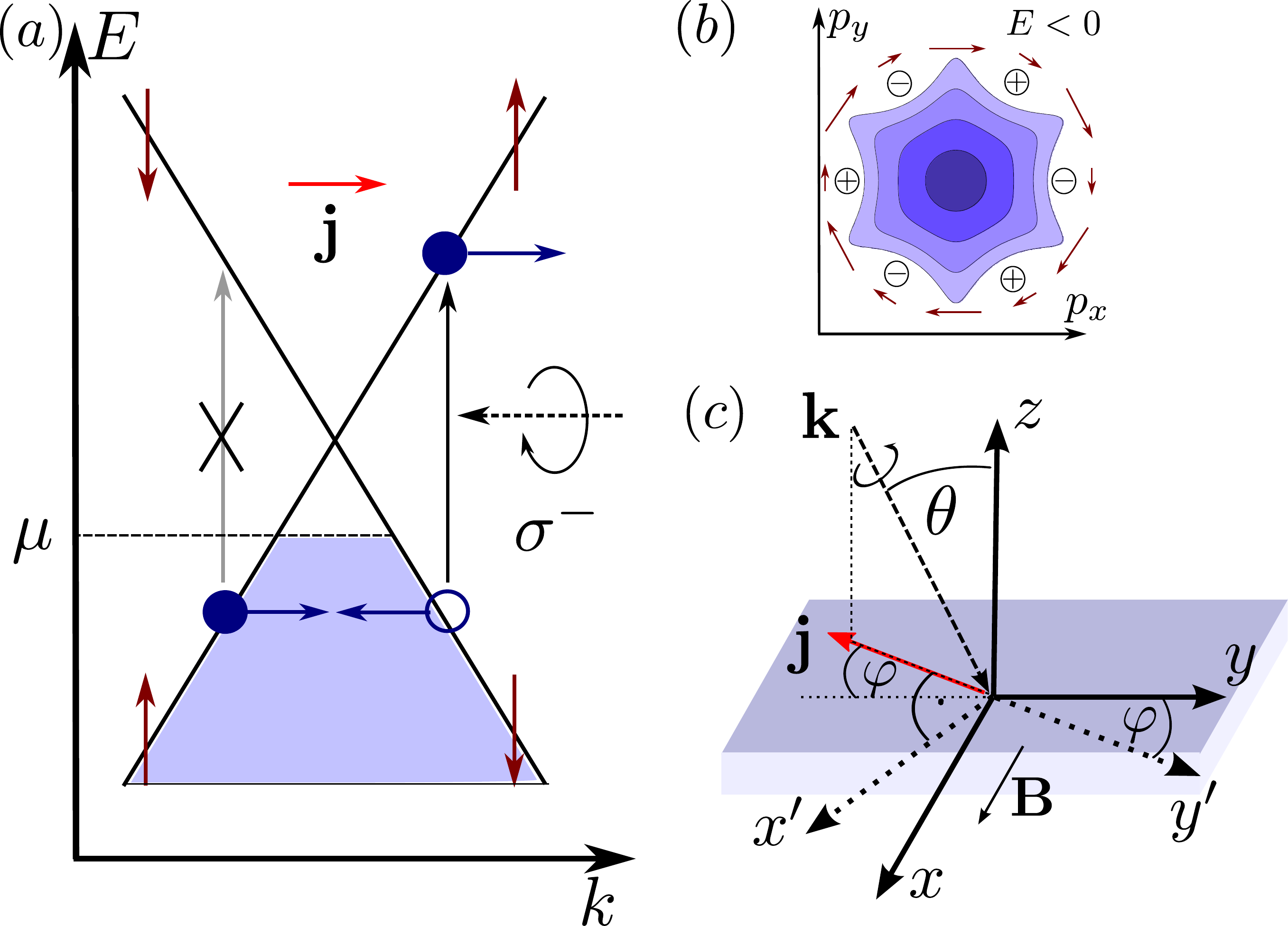}
 \caption{(color online) (a) Basic idea for the generation of a helicity-dependent photocurrent. Dark red arrows indicate spin direction, filled (empty) circles indicate electrons (holes). Circularly polarized light induces spin-dependent transitions, exciting electrons asymmetrically in $k$-space. (b) Illustration of the warping effect on the Fermi surface. Lighter colors/shades correspond to larger absolute values of energy. Dark red arrows indicate spins, circles with $+(-)$ indicate positive (negative) \mbox{spin $z$-component}. (c) Illustration of the orientation of the surface, the direction of the incident light and the resulting main contribution to the current. }
 \label{fig:schematic}
\end{figure}

The surface states also exhibit exotic optical properties. Gapped
surface states are predicted to cause giant Kerr and Faraday rotations
of polarized light.\cite{macdonald2010} The helical nature of 
surface states is expected to make their photocurrent
response to electromagnetic radiation rather unique.\cite{raghu2010,hosur2011} As illustrated in
Fig.~\ref{fig:schematic}(a), due to spin-selection rules, one might expect that circularly polarized light excites the surface states anisotropically around the Fermi surface, thus inducing a
$dc$ electric current. Such photocurrents constitute an interesting
probe of surface states in TIs (see, e.g.,
Ref. \onlinecite{mciver2011}). An important motivation for investigating the
photoresponse of TIs is that helicity-dependent currents are expected
to emerge solely due to the surface, as they require the breaking of
inversion symmetry precluding a second-order photogalvanic effect.\cite{hosur2011} Also, bulk electronic states in TIs consist of
Kramers pairs, which rules out a helicity dependence of the photocurrent
as well.\cite{mciver2011}

Here we investigate TI surface photocurrents within a minimal model of
the surface states motivated by
$\textnormal{Bi}_{2}\textnormal{Se}_{3}$, and obtain rather surprising results. The incident light couples
to the surface electrons in two ways: through orbital minimal coupling  (Peierls substitution), $\bm{p}\rightarrow \bm{p}-e\bm{A}$, and through
the Zeeman energy. Usually, the orbital effect is expected to dominate strongly, and the Zeeman coupling is neglected. Here, however, we find
that the orbital component of the photocurrent vanishes for the simplest model of a
perfect Dirac cone, as all surface-electron spins lie in the plane of
their motion, and light excites carriers isotropically around the
Dirac cone. The orbital coupling can induce photocurrents only when perturbations of
the ideal Dirac cone are included: hexagonal warping of the
Fermi surface;\cite{fu2009,wzhang2010,wang2011} an external magnetic
field; and a momentum-dependent correction to the Fermi
velocity.\cite{fu2009,wang2011,basak2011} Since these perturbations
are quite small, we also include the Zeeman coupling to the incident light in our analysis. 
Surprisingly, the Zeeman coupling is responsible for the dominant contribution to the surface photocurrent
response, which we find to be
helicity independent, linear in the Zeeman coupling, and to flow against
the direction of propagation of the light. The helicity-dependent
photocurrent, suggested by the simple mechanism illustrated in
Fig.~\ref{fig:schematic}(a), is found to be very small, i.e.,
quadratic in the Zeeman coupling.


{\em Model.---}We perform our analysis within a minimal model of a
TI surface. Let the surface lie in the $xy$-plane, with radiation incident at an arbitrary angle as illustrated in Fig.~\ref{fig:schematic}.
We assume that the energy $\hbar \omega$ of the incident radiation is
such that the excitation takes place solely within the Dirac cone
located within the bulk band gap ($E_{\textnormal{gap}}=0.3$ eV for
$\textnormal{Bi}_{2}\textnormal{Se}_{3}$\cite{xia2009,zhang2009}). As
usual, we neglect the momentum change in the optical
transitions. Motivated by the $(111)$ surface of $\textnormal{Bi}_{2}\textnormal{Se}_{3}$, we consider the model Hamiltonian
\begin{align}
H =  v(p_{x}\sigma_{y}-p_{y}\sigma_{x})+\frac{\lambda}{2}(p_+^{3}+p_-^{3})\sigma_{z}-g\mu_{B}B\sigma_{x},
\label{eq:hamiltonian}
\end{align}
which includes cubic warping $\lambda$,\cite{fu2009,wzhang2010,wang2011} a correction to the Fermi velocity $\Lambda$,\cite{fu2009,basak2011} and an external magnetic field $B$ in the $x$-direction.
Here $v=v_{F}+\Lambda p^2$ with the Fermi velocity $v_F$, $g$ denotes the g-factor, $\mu_B$ is the Bohr magneton, \mbox{$p_\pm=p_x\pm i p_y=p e^{\pm i \phi}$}, and $\sigma_\pm=\sigma_x\pm i \sigma_y$.\footnote{For quantitative estimates, we use $\lambda=50.1$~eV~$\text{\r{A}}^3/\hbar^3$,\cite{liu2010} and $\Lambda \sim 10-100$~eV~$\text{\r{A}}^3/\hbar^3$, which we deduced roughly using characteristic energy and length scales.} This Hamiltonian is partice-hole symmetric with $H\vert\bm{p},\pm\rangle=\pm E\vert\bm{p},\pm\rangle$. We consider circularly polarized light incident onto the sample at an azimuthal angle $\varphi$ from the negative $y$-axis and at a polar angle $\theta$ from the positive $z$-axis (cf. Fig.~\ref{fig:schematic}). For $\varphi=0$ and left-circular polarization, the vector potential is given by
 \begin{equation}
 \bm{{A}}(t)  =  A_{0}[\cos(\bm{k}\bm{r}-\omega t)\bm{\hat{x}}-\sin(\bm{k}\bm{r}-\omega t)(\bm{\hat{x}}\times \bm{\hat{k}})].
 \label{eq:vecpot}
 \end{equation}
While we focus on the response to circularly polarized light, we find that the surface also exhibits a photocurrent for linearly polarized radiation.
For $\varphi=0$, the direction of propagation $\bm{k}$ lies in the $yz$-plane at an angle $\theta$ from the positive $z$-axis, such that $\bm{\hat{k}}=\sin\theta\,\bm{\hat{y}}-\cos\theta\,\bm{\hat{z}}$ with $\theta\,\epsilon\,[0,\pi/2]$ (see Fig.\ref{fig:schematic}).

We include both orbital and Zeeman coupling of the light to the surface electrons in the coupling Hamiltonian
\begin{equation}
 H'=-e\bm{v_p}\cdot \bm{A}-g_s\mu_B(\nabla \times \bm{A})\cdot \hat{\boldsymbol{\sigma}},
 \label{eq:Hintgeneral}
\end{equation} 
where $m$ is the electron's mass, $\hat{\boldsymbol{\sigma}}$ the vector of Pauli matrices, and $\bm{v_p} = \partial{H}/\partial{\bm{p}}$. 
$H'$ encodes both photon absorption and emission processes, corresponding to creation or recombination of an electron-hole pair, respectively. It is, therefore, useful to write $H'=H'_+ +H'_-$ with $H'_{\pm}\propto A_{\pm}=A_0 e^{\pm i \omega t}$ describing the emission and absorption of a photon.

The relative magnitude of Zeeman and orbital coupling can be estimated as $\sim \hbar k/m v_F= \hbar \omega/m v_F c$. For a photon energy of $\hbar\omega=0.1$~eV and a Fermi velocity of $v_F=5\cdot10^{5}$~m/s,\cite{liu2010} this ratio is of order $\sim 10^{-5}$. Even though the orbital coupling clearly provides the dominant excitation process, we find that it produces no net photocurrent without an applied magnetic field. For this reason, we include the Zeeman coupling and find that the leading contribution to the current for realistic values of the magnetic field is given by an interference term between Zeeman and orbital coupling.


{\em Photocurrents.---}For a particle-hole symmetric system without spin degeneracy the current density in two dimensions can be written as 
\begin{equation}
 \bm{j}=-2e \sum_{\bm{p}} \bm{v}_{\bm{p},+} (n^{}_{\bm{p},+}-n_{\bm{p},+}^0),
\end{equation}
where the sum is over positive-energy states, $n^{}_{\bm{p},+}$ ($n_{\bm{p},+}^0$) is the distribution function (equilibrium distribution function) of momentum state $\bm{p}$ in the positive energy band, $\bm{v}_{\bm{p},+}$ is the velocity of a particle in state $\bm{p}$ in the positive energy band, and the factor of 2 accounts for particle-hole symmetry. Assuming that momentum relaxation occurs on a much faster timescale than energy relaxation, the steady-state solution of the Boltzmann equation in relaxation time approximation gives
\begin{equation}
 n^{}_{\bm{p},+}-n_{\bm{p},+}^0=\tau_p \,\Gamma_{\vert \bm{p},-\rangle \rightarrow \vert \bm{p},+\rangle}(n^{0}_{\bm{p},-}-n_{\bm{p},+}^0),
\label{eq:beq}
\end{equation} 
where $\tau_p$ is the momentum relaxation time and $\Gamma$ is the transition rate from state $\vert \bm{p},-\rangle$ in the lower band to state $\vert \bm{p},+\rangle$ in the upper band which can be calculated using Fermi's Golden Rule. For $T=0$, $(n^{0}_{\bm{p},-}-n_{\bm{p},+}^0)$ in Eq.~\ref{eq:beq} is only nonzero if the chemical potential lies between the energies of the two states participating in the transition (cf. Fig.~\ref{fig:schematic}). This gives a condition for the minimum photon energy required to induce transitions.

Assuming $\hbar \omega/2 > \vert \mu \vert$ for the photon energy, the current density becomes
\begin{equation}
 \bm{j}=-\frac{4\pi e \tau_p}{\hbar} \sum_{\bm{p}} \bm{v}_{\bm{p},+}\vert \langle\bm{p},+\vert H'_-\vert\bm{p},-\rangle \vert ^2 \delta(2E-\hbar \omega),
\label{eq:curr1}
\end{equation} 
where we used that $E_{\bm{p},+}=-E_{\bm{p},-}=E$ due to particle-hole symmetry.

The calculation of the integrand in Eq. (\ref{eq:curr1}) is expedited by using projection operators $\hat{P}_{\pm}$ onto the two bands,
\begin{equation}
-\bm{v}_{\bm{p},+}\vert \langle{\bm{p},+}\vert H'_-\vert\bm{p},-\rangle \vert ^2=\tr \left[\hat{P}_- H'_+ \hat{P}_+ H'_- \hat{P}_- \frac{\partial H}{\partial \bm{p}} \right],
\label{eq:integrand}
\end{equation} 
where we used that $\bm{v}_{\bm{p},+}=-\bm{v}_{\bm{p},-}$, that the expectation value of an operator can be written as 
\begin{equation}
 \langle \bm{p},\pm \vert \hat{O}\vert \bm{p},\pm \rangle=\tr[\hat{P}_\pm\hat{O}]=\tr\left[\left(\frac{\bm{1}}{2}\pm\frac{H}{2E} \right) \hat{O}\right],
\label{eq:expecval}
\end{equation} 
and that the transition $\vert \bm{p},+ \rangle \rightarrow \vert \bm{p},- \rangle$ ($\vert \bm{p},- \rangle \rightarrow \vert \bm{p},+ \rangle$) happens via emission (absorption) of a photon and is therefore mediated by the coupling involving $A_+$ ($A_-$).

In addition to being careful with the definition of the velocity, we must also make sure that the argument of the $\delta$-function in Eq.~\eqref{eq:curr1} contains the perturbations of the pure Dirac spectrum. To expand the argument of the $\delta$-function, we use $\delta(2E-\hbar \omega)=4E \int\limits_{-\infty}^{\infty} \frac{d\alpha}{2\pi} e^{i\alpha(4E^2-\hbar^2 \omega^2)}$. After expanding in the parameters $\lambda$, $\Lambda$, and $B$ to first order in $\lambda$ and $\Lambda$ and second order in $B$ and performing the angular integral in Eq.~\eqref{eq:curr1}, we find (see App.\ \ref{appA})
\begin{align}
 \bm{j}\simeq &\frac{4\pi e \tau_p}{\hbar}\int d\alpha\int\frac{dp\;p}{(2\pi\hbar)^{2}}\Big[\Xi_p^{(0)}+\alpha \Xi_p^{(1)}+\alpha^2 \Xi_p^{(2)}\Big]\nonumber\\
&\times e^{i\alpha [4(pv_F)^2-(\hbar \omega)^2]},
\label{eq:curr2}
\end{align} 
where $\Xi_p^{(i)}$ are functions of momentum and $\lambda$, $\Lambda$, and $B$. The integral over $\alpha$ is simplified by writing $\alpha\rightarrow -i\frac{\partial}{\partial(\hbar\omega)^2}$ for the factors of $\alpha$ in the brackets, and then first carrying out the $\alpha$-integral. The remaining integrals can now be easily done, since the complicated angular dependence of the eigenstates has been eliminated.


\renewcommand{\arraystretch}{1.4}
\begin{table}
\begin{center}
\resizebox{7.0cm}{!} {
  \begin{tabular}{ l |c| c | c }
$j^{(X)}_{\textnormal{hd}}$& prefactor& $x'$ & $y'$\\
    \hline \hline
    $0$ & $\frac{C}{4}\bar{v}_Z^2\sin\theta$&$1$ & ---\\ \hline
    $\lambda$&--- & --- & --- \\ \hline

$\Lambda$ & $-\frac{C}{16} \bar{v}_Z^2 \bar{\Lambda} \sin\theta$& $1$ & ---\\ \hline

    $B_1$ & $\frac{9 C}{32}\bar{v}_Z\bar{\lambda}\bar{B}\sin(2\theta)$&$\cos(2\varphi)$  & $\sin(2\varphi)$ \\ 
$B_2$& $\frac{3 C}{4}\bar{v}_Z^2\bar{\lambda}\bar{B}^2 \cos\theta$&$\cos\varphi$&$\sin\varphi$\\\hline

$\Lambda B_1$&$\frac{C}{4} \bar{v}_Z \bar{B} \bar{\Lambda}$ &$\cos\varphi$&$\cos^2\theta\sin\varphi$\\

$\Lambda B_2$&$\frac{15 C}{8} \bar{\lambda} \bar{B}^2 \bar{\Lambda} \cos\theta$ &$\cos\varphi$ &$\sin\varphi$\\

$\Lambda B_3$& $-\frac{27 C}{64} \bar{v}_Z  \bar{\lambda} \bar{B} \bar{\Lambda} \sin(2\theta)$&$\cos(2\varphi)$&$\sin(2\varphi)$\\

$\Lambda B_4$& $-\frac{C}{4}\bar{v}_Z^2 \bar{B}^2 \bar{\Lambda} \sin\theta$&$2+\cos(2\varphi)$&$\sin(2\varphi)$\\

$\Lambda B_5$&$-\frac{27 C}{16} \bar{v}_Z^2  \bar{\lambda} \bar{B}^2 \bar{\Lambda} \cos\theta$ &$\cos\varphi$&$\sin\varphi$\\ \hline
  \end{tabular}
}
\end{center}
\caption{Helicity-dependent corrections to the current induced by various perturbations. $\bar{v}_Z=v_Z/v_F\sim 10^{-5}$, $\bar{\lambda}=\lambda (\hbar\omega)^2/(v_F^3)\sim 10^{-2}$, $\bar{\Lambda}=\Lambda (\hbar\omega)^2/(v_F^3)\sim 10^{-3}$,\cite{Note1} and $\bar{B}=g\mu_B B/(\hbar\omega)\sim 10^{-4}B/$T are dimensionless parameters and $C$ is given in the text.}
\label{t:hd}
\end{table} 

\begin{table}
\begin{center}
\resizebox{8.5cm}{!} {
 \begin{tabular}{ l | c| c | c }
$j^{(X)}_{\textnormal{hi}}$&prefactor& $x'$ & $y'$\\
    \hline \hline
    $0$ & $-\frac{C}{4}\bar{v}_Z\sin\theta$& --- & $1$\\ \hline
    $\lambda$ & ---& --- & --- \\ \hline

$\Lambda$ & ---& --- & --- \\ \hline

$B_1$& $\frac{3C}{4}\bar{v}_Z  \bar{\lambda} \bar{B}^2\cos\theta$&$\sin\varphi$&$-\cos\varphi$ \\ 

$B_2$& $\frac{3 C}{32}\bar{v}_Z^2  \bar{\lambda} \bar{B}\sin(2\theta)$& $-\sin(2\varphi)$ & $\cos(2\varphi)$\\ \hline

$\Lambda B_1$& $\frac{C}{16}\bar{B}\bar{\Lambda}$&$-(3+\cos^2\theta)\sin\varphi$ &$(1+3\cos^2\theta)\cos\varphi$\\

$\Lambda B_2$&$\frac{C}{16}\bar{v}_Z^2 \bar{B}\bar{\Lambda}\sin^2\theta$ &$\sin\varphi$&$\cos\varphi$\\

$\Lambda B_3$&$\frac{33 C}{16}\bar{v}_Z \bar{\lambda} \bar{B}^2 \bar{\Lambda}\cos\theta$ &$-\sin\varphi$&$\cos\varphi$\\ 

$\Lambda B_4$& $\frac{17 C}{128}\bar{v}_Z^2 \bar{\lambda} \bar{B}\bar{\Lambda}\sin(2\theta)$&$\sin(2\varphi)$&$-\cos(2\varphi)$\\\hline
  \end{tabular}}
\end{center}
\caption{Helicity-independent corrections to the current induced by various perturbations. Parameters as in Tab.~\ref{t:hd}.}
\label{t:hi}
\end{table}

{\em Results for ideal Dirac spectrum.---}For the simplest model of a perfect Dirac cone without external fields, i.e., $\lambda=\Lambda=B=0$, and oblique incidence, we find that no net charge current is induced by pure orbital coupling. Although the coupling between vector potential and electron momentum leads to much larger excitation rates, the transitions take place isotropically around the Dirac cone and no net charge current is induced. However, including the small coupling between vector potential and electron spin, we find that currents are generated perpendicular and parallel to the plane of incidence with different polarization dependencies. We find that the current is given by
\begin{align}
 \bm{j}^{(0)}=&-\frac{C}{4}\bar{v}_Z  \sin\theta\left( \bm{\hat{y}}' -\bar{v}_Z \bm{\hat{x}}'\right),
\label{eq:j0}
\end{align} 
where $C=\frac{e^3E_0^2v_F\tau_p}{2 \omega\hbar^2}$, $\bar{v}_Z=\frac{g_s \hbar \omega}{2 m c v_F}$, and $\bm{\hat{x}}'$, $\bm{\hat{y}}'$ define a rotated coordinate system such that $\bm{A}$ is incident in the $y'z$-plane. $j^{(0)}_{y'}$ is helicity independent,  results from an interference effect between orbital and Zeeman coupling, and can also be induced by light which is linearly polarized perpendicular to the plane of incidence (S-polarized) (see App.\ \ref{appB}). The smaller current component $j^{(0)}_{x'}$ is helicity dependent, i.e., it changes sign for $\omega\rightarrow -\omega$, results from pure Zeeman coupling, and will not be present for linearly polarized light.

The result given by Eq.~\eqref{eq:j0} could have been anticipated by symmetry arguments. 
The (111) surface of $\textnormal{Bi}_{2}\textnormal{Se}_{3}$ has, among others, a mirror axis along the $y$-direction.\cite{fu2009} When the light is incident with $\varphi=0$, the helicity of the vector potential changes sign under this mirror transformation, the current in the $x$-direction changes sign, and the current in the $y$-direction remains invariant. Thus, helicity-dependent currents are only allowed in the $x$-direction while helicity-independent currents must flow in the $y$-direction. Since the system is rotationally symmetric for $\lambda=B=0$, the only directionality is provided by the vector potential and the currents will rotate accordingly. In addition, the interaction Hamiltonian for $\varphi=0$ is given by
\begin{align}
 H'_-\sim v_F\left(\sigma_y-i\cos\theta\sigma_x\right)-v_Z \left(\sigma_x+i\cos\theta\sigma_y+i\sin\theta\sigma_z\right).
\end{align} 
For normal incidence ($\theta=0$), the orbital coupling is proportional to the spin raising operator in the $z$-direction, $\sigma_+=\sigma_x+i\sigma_y$, while for oblique incidence it involves a sum of spin raising and lowering operators in the $z$-direction because $\sigma_y-i\cos\theta\sigma_x\sim\sigma_+(1+\cos\theta)-\sigma_-(1-\cos\theta)$. Since for $\lambda=0$ all spins lie in plane, electrons are excited isotropically around the Dirac cone and the orbital coupling by itself cannot generate a net current. For normal incidence the same argument even excludes currents induced by the Zeeman coupling. For oblique incidence, however, the Zeeman coupling involves, through $\sigma_x+i\sin\theta \sigma_z$, a sum of spin raising and lowering operators in the $y$-direction. Since the lowering operator has the larger coefficient, the Zeeman coupling will preferably excite spins with momentum in the negative $x$-direction generating a current in the $x$-direction (cf. Fig.~\ref{fig:schematic}). Similarly, we find that in the interference term the spin lowering operator in the $x$-direction $\sigma_z + i \sigma_y$ dominates, preferably exciting electrons with momentum in the positive $y$-direction. Interference between orbital and Zeeman couling thus leads to a current in the negative $y$-direction. 

Quantitatively, we estimate from Eq.~\eqref{eq:j0} $j^{(0)}_{y'}\sim 10$~$\mu$A/m and $j^{(0)}_{x'}\sim 1$~nA/m for the current densities parallel and perpendicular to the plane of incidence, respectively, using a laser power of $1$ W/mm${}^2$ as well as the parameters $g=g_s=1$ and $v_Z=\frac{g_s\hbar\omega}{2mc}=29$~m/s.  
For normal incidence the current vanishes, and for oblique incidence the dominant response is in the negative $y'$-direction since $\bar{v}_Z \sim10^{-5}$ (cf. Fig.~\ref{fig:schematic}). 

\begin{figure}[t]
\begin{minipage}{4cm}
	 \centering
 \includegraphics[width=4.2cm]{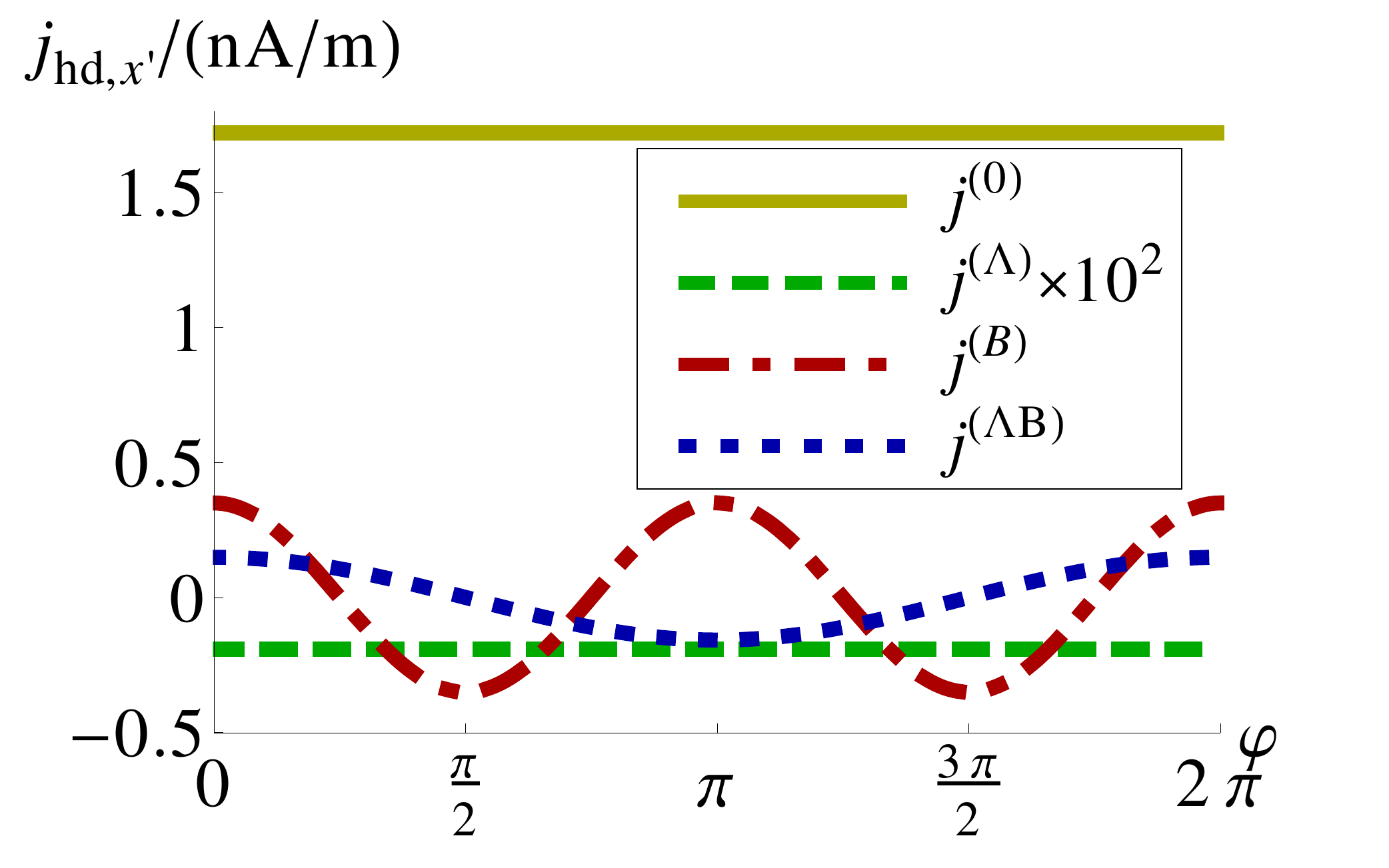}
\end{minipage}
\hfill
\begin{minipage}{4cm}
    \centering
	\includegraphics[width=4.2cm]{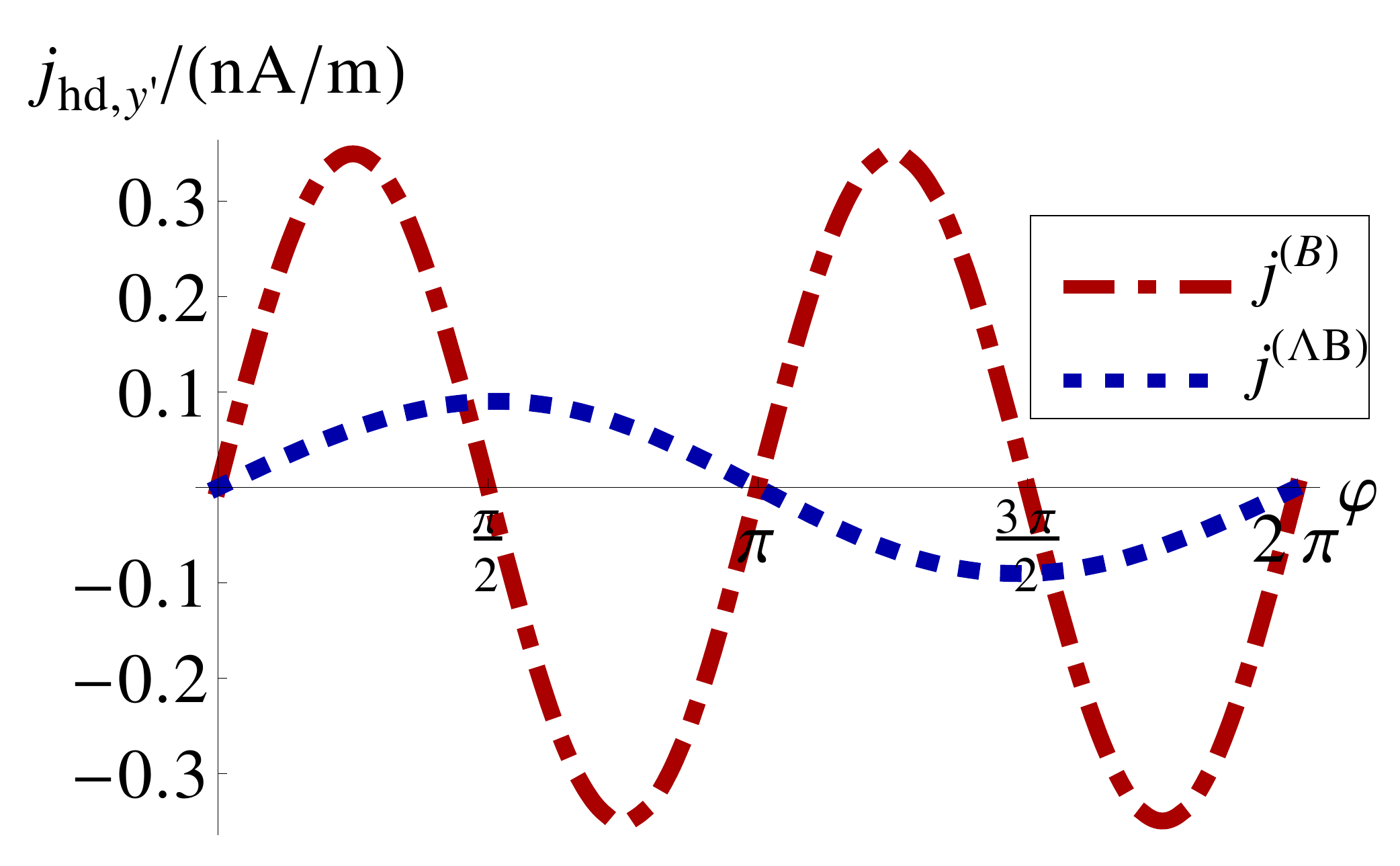}
\end{minipage}
 \begin{minipage}{4cm}
	 \centering
 \includegraphics[width=4.2cm]{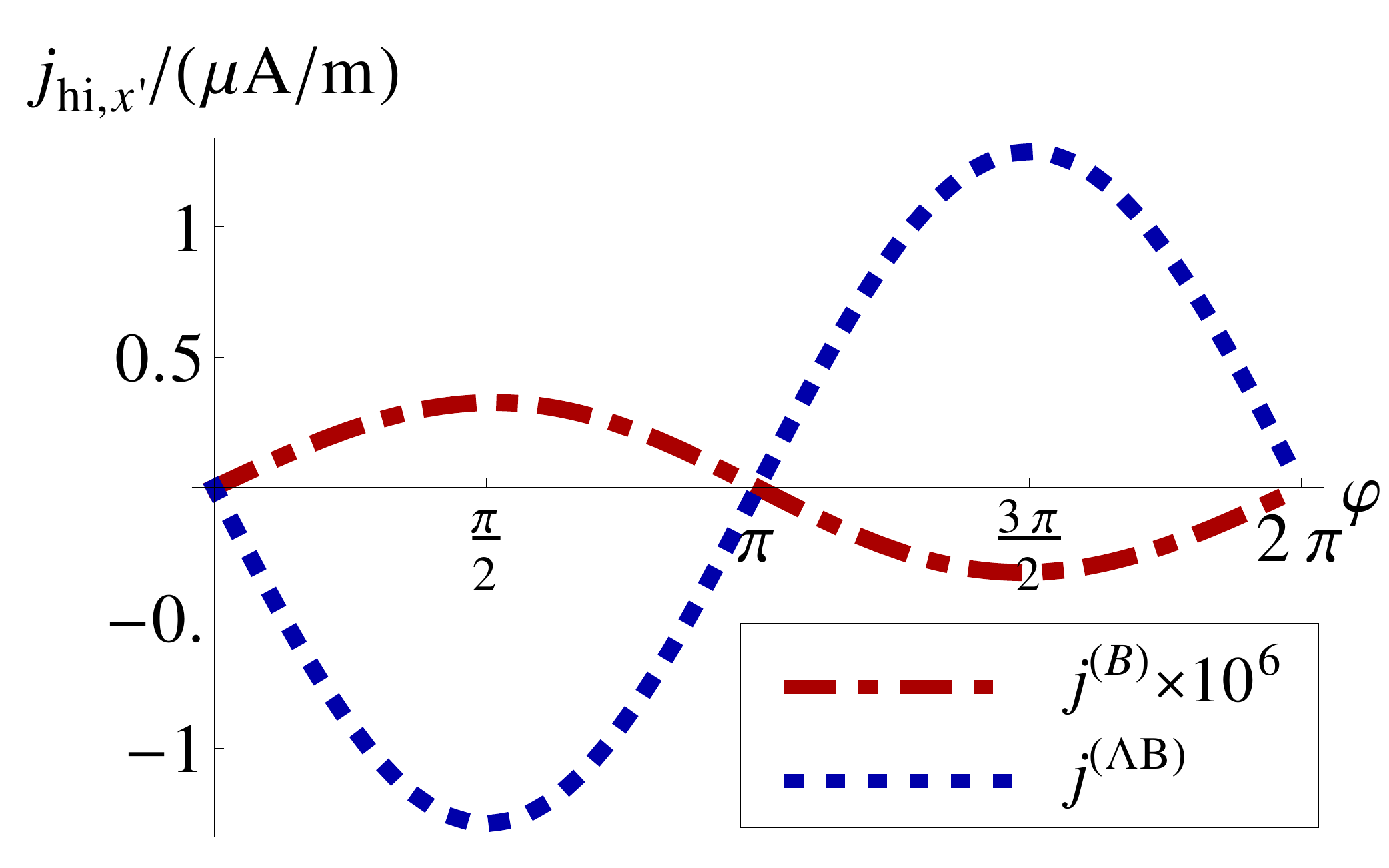}
\end{minipage}
\hfill
\begin{minipage}{4cm}
    \centering
	\includegraphics[width=4.2cm]{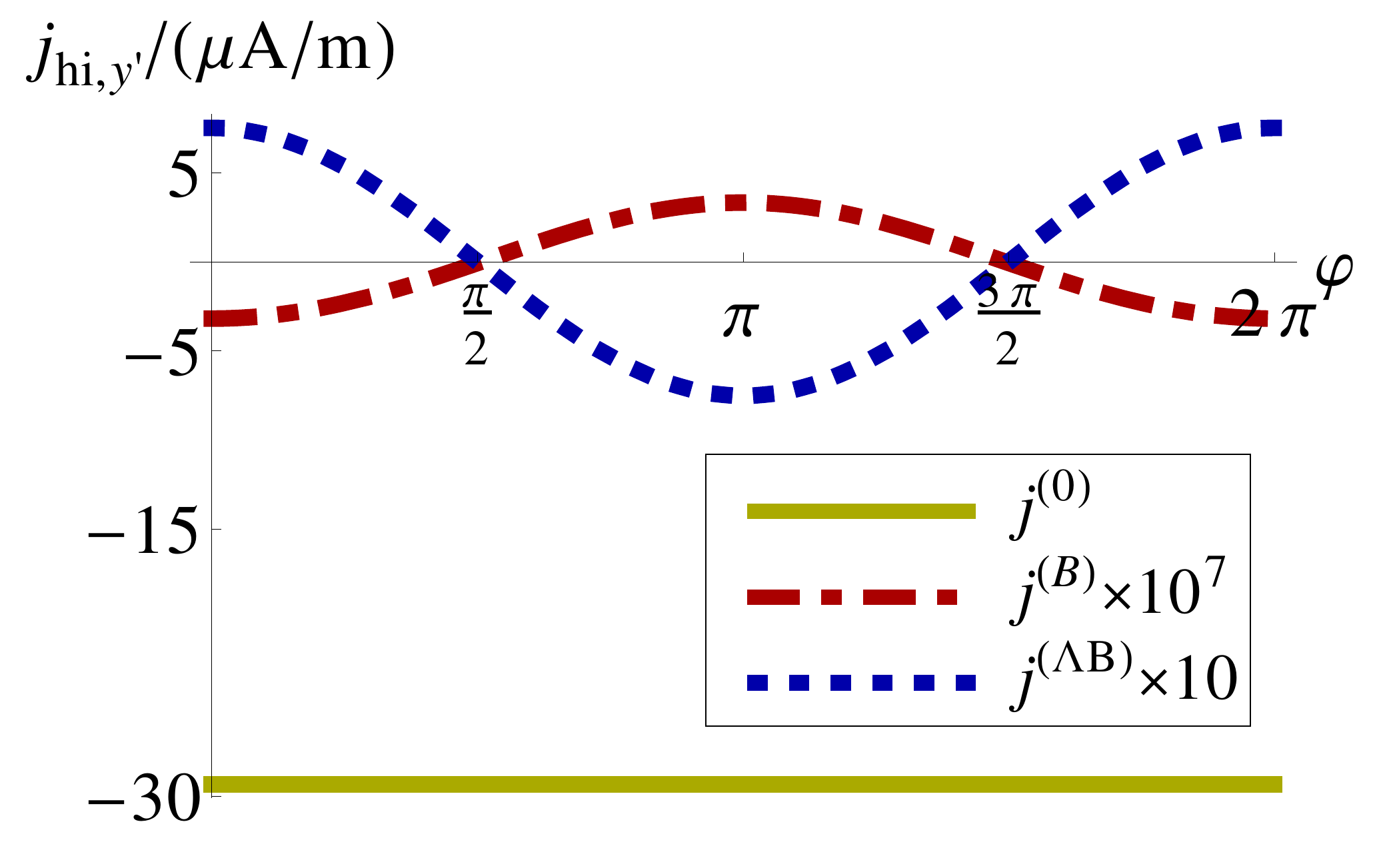}
\end{minipage}
\caption{Current as a function of polar angle $\varphi$ of vector potential. hd (hi): helicity dependent (independent). $\theta=0.98$ and the other parameters as in text.}
\label{fig:plots}
\end{figure}

{\em Results for more realistic dispersion.---}When we include deviations from the perfect Dirac cone in Eq.~\eqref{eq:hamiltonian}, there are additional helicity-dependent and independent contributions to the photocurrent, as listed in Tabs.~\ref{t:hd} and \ref{t:hi}. 
Without an external magnetic field, the leading correction to the helicity-dependent current is also in the $x'$-direction perpendicular to the plane of incidence. It is given by $j^{(\Lambda)}_{\textnormal{hd},x'}$ (see Tab.~\ref{t:hd}) with $j^{(\Lambda)}_{\textnormal{hd},x'}/j^{(0)}_{x'}\sim \bar{\Lambda}\sim 10^{-3}$ and results from pure Zeeman coupling. There are no helicity-independent corrections to the photocurrent $j^{(0)}_{y'}$, so while the leading response in the $x'$-($y'$-)direction is helicity dependent (independent), the overall leading response is helicity independent and parallel to the plane of incidence.

Including an external magnetic field of strength \mbox{$B=1$}~T, the leading helicity-dependent correction is $\bm{j}^{(B_1)}_{\textnormal{hd}}$ with components parallel and perpendicular to the plane of incidence. The relative magnitude is $j^{(B_1)}_{\textnormal{hd}}/j^{(0)}_{x'}\sim \bar{\lambda}\bar{B}/\bar{v}_Z\sim10^{-1}$ and results from interference between orbital and Zeeman coupling. The leading helicity-independent correction is $\bm{j}^{(\Lambda B_1)}_{\textnormal{hi}}$ also with components parallel and perpendicular to the plane of incidence. This results solely from orbital coupling and has a relative magnitude of $j^{(\Lambda B_1)}_{\textnormal{hi}}/j^{(0)}_{y'}\sim \bar{B}\bar{\Lambda}/\bar{v}_Z\sim 10^{-2}$. If the plane of incidence does not coincide with the $yz$-plane (cf. Fig.~\ref{fig:schematic}), the dominant response both parallel and perpendicular to the plane of incidence is helicity independent with the relative magnitude of the perpendicular component $j^{(0)}_{\textnormal{hd},x'}/j^{(\Lambda B_1)}_{\textnormal{hi},x'}\sim \bar{v}_Z^2/\bar{B}\bar{\Lambda}\sim 10^{-3}$.

For normal incidence, i.e., $\theta=\varphi=0$, $\bm{j}^{(0)}$ vanishes and there is no photocurrent, helicity-dependent or independent, in the absence of an external magnetic field. Since $j^{(0)}_{x'}$ even vanishes when including $B$, the helicity-dependent photocurrent for normal incidence is always significantly smaller than for oblique incidence. 
The leading contributions are $j^{(\Lambda B_1)}_{\textnormal{hd},x}$ and $j^{(\Lambda B_2)}_{\textnormal{hd},x}$ parallel to the magnetic field with $j^{(\Lambda B_1)}_{\textnormal{hd},x}/j^{(0)}_{x'}\sim \bar{B}\bar{\Lambda}/\bar{v}_Z\sim 10^{-2}$ and $j^{(\Lambda B_2)}_{\textnormal{hd},x}/j^{(0)}_{x'}\sim \bar{\lambda}\bar{B}^2\bar{\Lambda}/\bar{v}_Z^2\sim10^{-3}$. $j^{(\Lambda B_2)}_{\textnormal{hd},x}$ is the only helicity-dependent contribution induced by pure orbital coupling. The leading helicity-independent contribution again is $\bm{j}^{(\Lambda B_1)}_{\textnormal{hi}}$ perpendicular to the magnetic field. For normal incidence including a magnetic field, the leading response in the $x$-direction is helicity dependent, while the $y$-direction response is helicity independent.

The different contributions to the current as listed in Tabs.~\ref{t:hd} and \ref{t:hi} depend strongly on the angle of incidence of the laser and thus the relative magnitude can change significantly with azimuthal angle $\varphi$ of the vector potential. The dependence of the current on $\varphi$ for $\theta=0.98$ is plotted in Fig.~\ref{fig:plots}. The response $\bm{j}^{(0)}$ is independent of the polar angle but the corrections from $B$ and $\Lambda$ show strong angular dependence. The dominant current in the $y'$-direction is not affected by changes in the azimuthal angle but the dominant current in the $x'$-direction changes significantly as mentioned above. The large helicity-independent current $j^{(\Lambda B_1)}_{\textnormal{hi},x'}$ vanishes for light incident in the $yz$-plane.


{\em Conclusion.---}Motivated by recent experiments, we studied photocurrents in topological insulators. Focusing on the photocurrent response of the topological surface states, we find that the dominant photogalvanic current induced by obliquely incident circularly polarized light is helicity independent and in the plane of incidence of the light. This contribution is the result of an interference effect between the orbital and the Zeeman coupling of the light to the surface electrons. The helicity-dependent photocurrent is found to be very small. Although pure orbital coupling is the dominant excitation process, it does not induce a net photogalvanic charge current originating from the surface states unless when including both band curvature and an in-plane magnetic field. Our results suggest that an understanding of the experiments may require one to extend the theory to include the bulk states, the photon drag effect, or thermoelectric effects originating from inhomogeneous laser excitation. 

{\em Acknowledgements.---}We thank D.\ Hsieh for discussions and acknowledge financial support through the Helmholtz Virtual Institute ``New states of matter and their excitations'' (Berlin) as well as DARPA, the IQIM, an NSF institute supported by the Moore foundation, and the Humboldt foundation (Pasadena).

\appendix
\onecolumngrid
\section{Calculational Details}
\label{appA}
When calculating the photocurrent density induced on the surface of a topological insulator given by (Eq.~\eqref{eq:curr1} in the main text),
\begin{equation}
 \bm{j}=-\frac{4\pi e \tau_p}{\hbar}\sum_{\bm{p}}\bm{v}_{\bm{p},+}\vert \langle\bm{p},+\vert H'_-\vert\bm{p},-\rangle \vert ^2 \delta(2E-\hbar \omega),
\label{eq:curr1app}
\end{equation} 
one needs to be careful to include the perturbations on the Dirac spectrum in the interaction matrix element, as well as in the velocity and the $\delta$-function. 

This section is organized as follows. First we will show how to calculate the integrand $\bm{v}_{\bm{p},+}\vert \langle\bm{p},+\vert H'_-\vert\bm{p},-\rangle \vert ^2$. Second, we will address how to treat the Delta function before calculating the entire integral in Eq.~\eqref{eq:curr1app}.

The calculation in detail of the integrand in Eq.~\eqref{eq:curr1app} proceeds as follows. Using projection operators onto the two bands defined as
\begin{equation}
 \hat{P}_\pm:=\vert\bm{p},\pm\rangle\langle \bm{p},\pm \vert=\frac{\bm{1}}{2}\pm\frac{H}{2E},
\label{eq:projection}
\end{equation} 
where $E_+$ is the full energy with all corrections to the perfect Dirac cone, the integrand can be expressed as
\begin{align}
-\bm{v}_{\bm{p},+}\vert \langle\bm{p},+\vert H'_-\vert\bm{p},-\rangle \vert ^2&=\langle \bm{p},-\vert H'_+\vert\bm{p},+\rangle\langle\bm{p},+\vert H'_-\vert \bm{p},-\rangle\langle \bm{p},-\vert \frac{\partial H}{\partial \bm{p}}\vert \bm{p},-\rangle =\tr \left[\hat{P}_- H'_+ \hat{P}_+ H'_- \hat{P}_- \frac{\partial H}{\partial \bm{p}} \right].
\label{eq:integrandapp}
\end{align} 
In the first step we used that the transition $\vert \bm{p},+ \rangle \rightarrow \vert \bm{p},- \rangle$ ($\vert \bm{p},- \rangle \rightarrow \vert \bm{p},+ \rangle$) happens via emission (absorption) of a photon and is therefore mediated by the coupling involving $A_+$ ($A_-$). The minus sign is included by taking the expectation value of the velocity operator with respect to the lower band. In the second step we used that the expectation value of an operator can be written as 
\begin{equation}
 \langle\bm{p},\pm\vert \hat{O}\vert\bm{p},\pm\rangle=\tr[\hat{P}_\pm\hat{O}]=\tr\left[\left(\frac{\bm{1}}{2}\pm\frac{H}{2E}\right) \hat{O}\right].
\label{eq:expecvalapp}
\end{equation} 

In order to calculate Eq.~\eqref{eq:integrandapp} explicitly, we need to compute the interaction Hamiltonian given by (Eq.~\eqref{eq:Hintgeneral} in the main text)
\begin{equation}
 H'=-e\bm{v}_{\bm{p},+}\cdot \bm{A}-g_s\mu_B(\nabla \times \bm{A})\cdot \hat{\boldsymbol{\sigma}}.
 \label{eq:Hintgeneralapp}
\end{equation} 
Since we would like to separate terms which create/annihilate a photon, i.e., write the interaction Hamiltonian in terms of $A_{\pm}$, we write the velocities as
\begin{align}
 \frac{\partial H}{\partial p_x}&=\frac{\partial H}{\partial p_+}\frac{\partial p_+}{\partial p_x}+\frac{\partial H}{\partial p_-}\frac{\partial p_-}{\partial p_x}= v_+ + v_-&
\text{and}&&
\frac{\partial H}{\partial p_y}&=\frac{\partial H}{\partial p_+}\frac{\partial p_+}{\partial p_y}+\frac{\partial H}{\partial p_-}\frac{\partial p_-}{\partial p_y}= i(v_+ - v_-).
\label{eq:velocity2}
\end{align} 
with $p_{\pm}=p_x \pm i p_y$ and $\partial H/ \partial p_{\pm} =v_{\pm}$.
The velocity operator then becomes
\begin{equation}
 \frac{\partial H}{\partial \bm{p}}=\frac{\partial H}{\partial p_x}\bm{\hat{x}}+\frac{\partial H}{\partial p_y}\bm{\hat{y}}=\left[v_+(\bm{\hat{x}}+i \bm{\hat{y}})+v_-(\bm{\hat{x}}- i\bm{\hat{y}})\right].
\label{eq:velocity}
\end{equation} 
Here $v_{\pm}$ contains all corrections to the perfect Dirac cone.
The vector potential for left circular polarization, $\bm{\hat{k}}=\sin\theta\,\bm{\hat{y}}-\cos\theta\,\bm{\hat{z}}$ with $\theta\,\epsilon\,[0,\pi/2]$, and $\varphi=0$ is given by
\begin{eqnarray}
\bm{{A}}(t) & = & A_{0}[\cos(\bm{k}\cdot\bm{r}-\omega t)\hat{{\bm{x}}}-\sin(\bm{k}\cdot\bm{r}-\omega t)(\bm{\hat{x}}\times\bm{\hat{k}})]\nonumber\\
 & = & \frac{{1}}{2}[A_+e^{-i\bm{k}\cdot\bm{r}}(\bm{\hat{x}}-i\cos(\theta)\bm{\hat{y}}-i\sin(\theta)\bm{\hat{z}})+A_-e^{i\bm{k}\cdot\bm{r}}(\bm{\hat{x}}+i\cos(\theta)\bm{\hat{y}}+i\sin(\theta)\bm{\hat{z}})].
\label{eq:vecpot2}
\end{eqnarray}
with $A_{\pm}=A_0e^{\pm i \omega t}$.
From this we can read that
\begin{eqnarray}
 A_x=\frac{1}{2} (A_++A_-)&\qquad\text{and}\qquad&A_y=\frac{1}{2i} \cos\theta (A_+-A_-).
\end{eqnarray} 
With these expressions the orbital part of the interaction Hamiltonian becomes
\begin{align}
 \frac{\partial H}{\partial p_x}A_x(t)+\frac{\partial H}{\partial p_y}A_y(t)&=\frac{1}{2}(v_++v_-)(A_++A_-)+\frac{i}{2}(v_+ - v_-)(-i \cos\theta)(A_+-A_-)\nonumber\\
&=\frac{1}{2}\big\{A_+\left[v_+(1+\cos\theta)+v_-(1-\cos\theta)\right]+A_-\left[v_+(1-\cos\theta)+v_-(1+\cos\theta)\right]\big\},
\label{eq:orbcoup}
\end{align}
with $H_{\pm}$ being the part proportional to $A_{\pm}$.
In order to calculate the Zeeman term in the interaction using Eq.~\eqref{eq:vecpot2} we can calculate
\begin{align}
 \nabla \times \bm{A}&=\frac{1}{2}\Big\{A_+\big[-i\bm{k}\times(\bm{\hat{x}}-i\cos(\theta)\bm{\hat{y}}-i\sin(\theta)\bm{\hat{z}})\big]e^{-i\bm{k}\cdot\bm{r}}+A_-\big[i\bm{k}\times(\bm{\hat{x}}+i\cos(\theta)\bm{\hat{y}}+i\sin(\theta)\bm{\hat{z}})\big]e^{i\bm{k}\cdot\bm{r}}\Big\}\nonumber\\
&=\frac{k}{2}\Big\{A_+\big[-\bm{\hat{x}}+i\cos(\theta)\bm{\hat{y}}+i\sin(\theta)\bm{\hat{z}}\big]e^{-i\bm{k}\cdot\bm{r}}+A_-\big[-\bm{\hat{x}}-i\cos(\theta)\bm{\hat{y}}-i\sin(\theta)\bm{\hat{z}}\big]e^{i\bm{k}\cdot\bm{r}}\Big\}\nonumber\\
&=-k\bm{A}.
\label{eq:curlA}
\end{align} 
Note that for right circularly polarized light we would get the opposite sign in the last line of Eq.~\eqref{eq:curlA},i.e. , $\nabla \times \bm{A}_{\textnormal{RCP}}=k\bm{A}_{\textnormal{RCP}}$. With Eq.~\eqref{eq:curlA} the Zeeman coupling can be written as
\begin{equation}
 (\nabla \times \bm{A})\cdot \hat{\boldsymbol{\sigma}}=-\frac{k}{2}\left[A_+(\sigma_x-i \cos\theta \sigma_y -i \sin\theta \sigma_z)+A_-(\sigma_x+i \cos\theta \sigma_y +i \sin\theta \sigma_z)\right].
\label{eq:spincoup}
\end{equation} 
Inserting Eqs.~\eqref{eq:orbcoup} and \eqref{eq:spincoup} into the interaction Hamiltonian given by Eq.~\eqref{eq:Hintgeneral} in the main text, we get
\begin{align}
 H'=&-\frac{e}{2}\Big\{A_+\left[v_+(1+\cos\theta)+v_-(1-\cos\theta)\right]+A_-\left[v_+(1-\cos\theta)+v_-(1+\cos\theta)\right]\nonumber\\
&-v_Z\big[A_+(\sigma_x-i \cos\theta \sigma_y -i \sin\theta \sigma_z)+A_-(\sigma_x+i \cos\theta \sigma_y +i \sin\theta \sigma_z)\big]\Big\},
\end{align}
where $v_Z=\frac{g_s\hbar k}{2m}$ is helicity dependent. The integrand can now be explicitly written as
\begin{align}
 -\bm{v}_{\bm{p},+}\vert \langle\bm{p},+\vert H'\vert\bm{p},-\rangle \vert ^2=&\frac{e^2}{4}\tr\Big\{\hat{P}_-A_+\big[v_+(1+\cos\theta)+v_-(1-\cos\theta)-v_Z(\sigma_x-i\cos\theta \sigma_y - i \sin\theta \sigma_z)\big]\nonumber\\
&\quad\quad\;\; \hat{P}_+ A_-\big[v_+(1-\cos\theta)+v_-(1+\cos\theta)-v_Z(\sigma_x+i\cos\theta \sigma_y + i \sin\theta \sigma_z)\big]\hat{P}_-\nonumber\\
&\quad\quad\quad \left[v_+(\bm{\hat{x}}+i \bm{\hat{y}})+v_-(\bm{\hat{x}}- i\bm{\hat{y}})\right]\Big\},
\label{eq:integrand2}
\end{align}
where the trace is understood to be over matrices only, not including the unit vectors. Eq.~\eqref{eq:integrand2} is exact and contains all contributions from perturbations on the perfect Dirac spectrum arising from the interaction matrix element and the velocity.

In addition to being careful with the definition of the velocity, we must also make sure the delta function expresses the perturbation on the Dirac spectrum. In order to Taylor-expand the argument of the $\delta$-function, we use
\begin{equation}
 \delta(2E-\hbar \omega)=4E \delta(4E^2-\hbar^2 \omega^2)=4E\int\limits_{-\infty}^{\infty} \frac{d\alpha}{2\pi} e^{i\alpha(4E^2-\hbar^2 \omega^2)}.
\label{eq:deltafct}
\end{equation}  
After expanding the entire integrand of the momentum integral in Eq.~\eqref{eq:curr1app}, i.e., Eq.~\eqref{eq:integrand2} and \eqref{eq:deltafct}, in the parameters $\lambda$, $\Lambda$, and $B$ to first order in $\lambda$ and $\Lambda$ and second order in $B$, and performing the angular integral, we find
\begin{align}
 \bm{j}\simeq &\frac{4\pi e \tau_p}{\hbar}\int d\alpha\int\frac{dp\;p}{(2\pi\hbar)^{2}}\Big[\Xi^{(0)}(p)+\alpha \Xi^{(1)}(p)+\alpha^2 \Xi^{(2)}(p)\Big] e^{i\alpha [4(pv_F)^2-(\hbar \omega)^2]},
\label{eq:curr2app}
\end{align} 
where $\Xi^{(i)}(p)$ are functions of momentum and contain the parameters $\lambda$, $\Lambda$, and $B$ to the desired order. The exlicit expressions for the $\Xi^{(i)}(p)$ are very long without giving any insight and will not be presented here. The integral over $\alpha$ is simplified by writing $\alpha\rightarrow -i\frac{\partial}{\partial(\hbar\omega)^2}$ for the factors of $\alpha$ in the brackets. Integrating each summand separately, the derivatives can be pulled in front of the integrals. The integration over $\alpha$ in combination with the exponential function can now be resubstituted by a $\delta$-function. Using $\delta(4(pv_F)^2-\xi)=\frac{1}{8v_F^2p}\delta\Big(p-\frac{\sqrt{\xi}}{2v_F}\Big)$ with $\xi=(\hbar\omega)^2$, Eq.~\eqref{eq:curr2app} becomes
\begin{align}
  \bm{j}&\simeq \frac{4e\pi}{\hbar}\frac{1}{(2\pi\hbar)^2}\tau_p\Big\{\int dp p\, \Xi^{(0)}(p) \frac{1}{8v_F^2p}\delta\Big(p-\frac{\sqrt{\xi}}{2v_F}\Big)+i \frac{\partial}{\partial \xi}\int dp p\, \Xi^{(1)}(p) \frac{1}{8v_F^2p}\delta\Big(p-\frac{\sqrt{\xi}}{2v_F}\Big)\nonumber\\
&\quad\quad\quad\quad\quad\quad\quad\quad-\frac{\partial^2}{\partial \xi^2}\int dp p\, \Xi^{(2)}(p) \frac{1}{8v_F^2p}\delta\Big(p-\frac{\sqrt{\xi}}{2v_F}\Big)\Big\}\nonumber\\
&= \frac{4e\pi \tau_p}{\hbar}\frac{1}{(2\pi\hbar)^2}\frac{1}{8v_F^2}\Big\{\Xi^{(0)}\bigg(\frac{\sqrt{\xi}}{2v_F}\bigg)+i \frac{\partial}{\partial \xi}\Xi^{(1)}\bigg(\frac{\sqrt{\xi}}{2v_F}\bigg)-\frac{\partial^2}{\partial \xi^2}\Xi^{(2)}\bigg(\frac{\sqrt{\xi}}{2v_F}\bigg)\Big\}.
\end{align} 
This can be easily evaluated leading to the results given in the main text.

\section{Linear Polarization}
\label{appB}

\begin{table}[t]
\begin{center}
\resizebox{8.5cm}{!} {
  \begin{tabular}{ l |c| c | c }
$j^{(X)}_{P}$& prefactor& $x'$ & $y'$\\
    \hline \hline
    $0$ & ---& --- & ---\\ \hline
    $\lambda$& $\frac{3C}{32} \bar{v}_Z \bar{\lambda}\cos\theta$ & $-\sin(3\varphi)$ & $\cos(3\varphi)$ \\ \hline

$\Lambda$ & $\frac{5C}{64} \bar{v}_Z \bar{\lambda}\bar{\Lambda}\cos\theta$& $\sin(3\varphi)$ & $-\cos(3\varphi)$\\ \hline

$B$ & $\frac{3 C}{4}\bar{v}_Z\bar{\lambda}\bar{B}^2 \cos\theta$&$\sin\varphi$  & --- \\\hline

$\Lambda B_1$&$\frac{C}{16} \bar{B} \bar{\Lambda}$ &$-\cos^2\theta\sin\varphi$&$3\cos^2\theta\cos\varphi$\\

$\Lambda B_2$&$\frac{C}{16}\bar{v}_Z^2 \bar{B}\bar{\Lambda}$ &$\sin\varphi$ &$\cos\varphi$\\

$\Lambda B_3$& $\frac{3 C}{16} \bar{v}_Z  \bar{\lambda} \bar{B}^2 \bar{\Lambda} \cos\theta$&$-9\sin\varphi+5\sin(3\varphi)$& $\cos\varphi[7-10\cos(2\varphi)]$\\ \hline
  \end{tabular}
}
\end{center}
\caption{P-polarization. $\bar{v}_Z=v_Z/v_F\sim 10^{-5}$, $\bar{\lambda}=\lambda (\hbar\omega)^2/(v_F^3)\sim 10^{-2}$, $\bar{\Lambda}=\Lambda (\hbar\omega)^2/(v_F^3)\sim 10^{-3}$, and $\bar{B}=g\mu_B B/(\hbar\omega)\sim 10^{-4}B/$T are dimensionless parameters and $C$ is given in the main text.}
\label{t:ppol}
\end{table} 

\begin{table}[t]
\begin{center}
\resizebox{9cm}{!} {
 \begin{tabular}{ l | c| c | c }
$j^{(X)}_{S}$&prefactor& $x'$ & $y'$\\
    \hline \hline
    $0$ & $-\frac{C}{4}\bar{v}_Z\sin\theta$& --- & $1$\\ \hline
    $\lambda$& $\frac{3C}{32} \bar{v}_Z \bar{\lambda}\cos\theta$ & $\sin(3\varphi)$ & $-\cos(3\varphi)$ \\ \hline

$\Lambda$ & $\frac{5C}{64} \bar{v}_Z \bar{\lambda}\bar{\Lambda}\cos\theta$& $-\sin(3\varphi)$ & $\cos(3\varphi)$ \\ \hline

$B_1$& $\frac{3C}{4}\bar{v}_Z  \bar{\lambda} \bar{B}^2\cos\theta$&---&$-\cos\varphi$ \\ 

$B_2$& $\frac{3 C}{32}\bar{v}_Z^2  \bar{\lambda} \bar{B}\sin(2\theta)$& $-\sin(2\varphi)$ & $\cos(2\varphi)$\\ \hline

$\Lambda B_1$& $\frac{C}{16}\bar{B}\bar{\Lambda}$&$-3\sin\varphi$ &$\cos\varphi$\\

$\Lambda B_2$&$-\frac{C}{16}\bar{v}_Z^2 \bar{B}\bar{\Lambda}\cos^2\theta$ &$\sin\varphi$&$\cos\varphi$\\

$\Lambda B_3$&$\frac{3 C}{16}\bar{v}_Z \bar{\lambda} \bar{B}^2 \bar{\Lambda}\cos\theta$ &$-[2\sin\varphi+5\sin(3\varphi)]$&$9\cos\varphi+5\cos(3\varphi)$\\ 

$\Lambda B_4$& $\frac{17 C}{128}\bar{v}_Z^2 \bar{\lambda} \bar{B}\bar{\Lambda}\sin(2\theta)$&$\sin(2\varphi)$&$-\cos(2\varphi)$\\\hline
  \end{tabular}}
\end{center}
\caption{S-polarized. Parameters as in Tab.~\ref{t:ppol}.}
\label{t:spol}
\end{table}

For completeness we also calculated the response for P- and S-linearly polarized light. For $\bm{\hat{k}}=\sin\theta\,\bm{\hat{y}}-\cos\theta\,\bm{\hat{z}}$ and $\phi=0$, the vector potentials for P- and S-polarized light are given by
\begin{equation}
\bm{A}_{\textnormal{P}}=A_0 \cos(\bm{k} \cdot \bm{r}-\omega t) [\cos\theta \,\bm{\hat{y}}+\sin\theta \,\bm{\hat{z}}]
\end{equation}
and
\begin{equation}
\bm{A}_{\textnormal{S}}=A_0 \cos(\bm{k} \cdot \bm{r}-\omega t) \,\bm{\hat{x}}.
\end{equation}
The calculation of the photocurrent proceeds as for circularly polarized light and the results are listed in Tabs.~\ref{t:ppol} and \ref{t:spol}. The resulting currents are, of course, helicity independent and the sum of the contributions from S- and P-polarized light add up to the helicity independent photocurrent induced by circular polarized light.
As mentioned in the main text, the overall largest contribution to the photocurrent, which is helicity independent and in the direction opposite to the direction of propagation of the incident light, can by induced by S-polarized light.
\twocolumngrid
\bibliographystyle{apsrev}

\onecolumngrid

\end{document}